\begin{filecontents}{leer.eps}
gsave
72 31 moveto
72 342 lineto
601 342 lineto
601 31 lineto
72 31 lineto
showpage
grestore
\documentclass[epj]{svjour}
\usepackage{graphics}
\begin{document}
\title{Anisotropic generalization of well-known solutions describing relativistic self-gravitating fluid systems: An
algorithm}
\titlerunning{Anisotropic generalization of well-known solutions}

\author{S. Thirukkanesh\inst{1} \and F. C. Ragel\inst{2} \and Ranjan Sharma\inst{3}\thanks{E-mail: rsharma@associates.iucaa.in} \and Shyam Das\inst{3} 
}
\authorrunning{Thirukkanesh {\em et al}}


\institute{Department of Mathematics, Eastern University, Chenkalady, Sri Lanka. \and Department of Physics, Eastern University, Chenkalady, Sri Lanka. \and Department of Physics, P. D. Women's College, Jalpaiguri 735101, West Bengal, India.}
\date{Received: date / Revised version: date}
%
\abstract{We present an algorithm to generalize a plethora of well-known solutions to Einstein field equations describing spherically symmetric relativistic fluid spheres by relaxing the pressure isotropy condition on the system. By suitably fixing the model parameters in our formulation, we generate closed-form solutions which may be treated as an anisotropic generalization of a large class of solutions describing isotropic fluid spheres. From the resultant solutions, a particular solution is taken up to show its physical acceptability. Making use of the current estimate of mass and radius of a known pulsar, the effects of anisotropic stress on the gross physical behaviour of a relativistic compact star is also highlighted.
} 
        
\maketitle

\section{\label{sec1} Introduction}
In relativistic astrophysics, there has been a growing interest in studying the physical behaviour of stellar objects composed of anisotropic fluid distributions, i.e., objects where the radial component of pressure $(p_r)$ is not equal to its transverse component $(p_t)$. A Newtonian approach is sufficient to study stellar structures in a comparatively low-density regime. However, in the case of compact stellar structures in the high-density regime, a general relativistic treatment is necessary, and the impact of anisotropy cannot be neglected while modelling such systems, see for example \cite{Herrera13,Herrera13b,Dev03,Shojai15} and references therein. Ruderman\cite{Ruderman} and Canuto\cite{Canuto} observed that material distribution in the highly dense core of a compact star might exhibit unequal stresses. Bowers and Liang\cite{Bowers} have extensively analyzed the sources of anisotropy at the stellar interior. 

Pressure anisotropy in compact star may arise due to various factors which include phase transitions, pion condensation\cite{Sokolov,Herrera89,Sawyer}, the existence of a solid core or presence of a type$-3A$ superfluid\cite{Kippen}, strong electromagnetic fields\cite{Weber,Perez,Usov}, slow rotation of fluids\cite{Herrera95}, etc. Ivanov\cite{Ivanov10} has pointed out that influences of shear and/or electromagnetic field on self-bound systems can be interpreted by incorporating a gross anisotropic parameter into the system of field equations. Self-bound systems composed of scalar fields, i.e., the `boson stars' are naturally anisotropic\cite{Schunck}. Similarly, wormholes\cite{Morris} and gravastars\cite{Cattoen,DeBend} are also structurally anisotropic systems. The shearing motion of the fluid is another source of anisotropy in self-gravitating objects\cite{Chan1,Chan2,Chane}. The origin and effects of local anisotropy on astrophysical objects have been studied in details in \cite{Chan6,Herrera5}. An exhaustive review of the subject may be found in \cite{Herrera6}.

The objective of the current investigation is to provide a new algorithm to generate anisotropic analogues of a large family of well-known solutions describing self-gravitating systems in equilibrium. An algorithm to generate anisotropic solutions from a seed isotropic solution was initiated by Chaisi and Maharaj\cite{Mosa06}.  Herrera {\em et al}\cite{Herrera08} had extended the Lake\cite{Lake03} algorithm to the case of locally anisotropic fluids to study spherically symmetric relativistic stars. Herrera and Barreto\cite{Herrera13b} had set up a general formalism to model relativistic polytropic stars with anisotropic pressure. In our formalism, we have shown that it is possible to generalize a large class of well-known exact isotropic stellar solutions by extending the models to the case of an anisotropic matter distribution. Most importantly, the resultant solution fulfils the criteria of physical acceptability.   

The paper has been organized as follows: In Section \ref{sec2}, the Einstein field equations for a static spherically symmetric anisotropic fluid distribution have been laid down. An equivalent form of the field equations has been obtained by making use of the Durgapal and Bannerji\cite{Durga} transformation equations. In Section \ref{sec3}, for a particular choice of the $g_{tt}$ component of the gravitational potential together with a prescribed form of the anisotropic parameter, a formalism to has been developed to generate analytic solutions in terms elementary functions. In Section \ref{sec4}, we have shown how a plethora of physically reasonable isotropic stellar solutions can be regained by suitable parametrization of our general class of solutions describing an anisotropic matter distribution. In Sections \ref{sec5} and \ref{sec6}, we have analyzed physical acceptability and implications of our class of solutions on the gross physical behaviour of relativistic compact stars. Some concluding remarks have been made in Section \ref{sec7}.

\section{\label{sec2} Einstein field equations:}
To describe the interior of a static and  spherically symmetric relativistic star, we write the line element in cordinates $(x^{a}) = (t,r,\theta,\phi)$ as 
\begin{equation}
\label{1} ds^{2} = -e^{2\nu(r)} dt^{2} + e^{2\lambda(r)} dr^{2} +
 r^{2}(d\theta^{2} + \sin^{2}{\theta} d\phi^{2}).
\end{equation}
For an anisotropic matter distribution, we choose the energy-momentum tensor in the form  
\begin{equation}
\label{2} T^i_{j}=\mbox{diag}(-\rho, p_r, p_t, p_t).
\end{equation}
The energy density $\rho$, the radial pressure $p_r$ and  the tangential pressure $p_t$ are measured relative to the comoving
fluid velocity $u^i = e^{-\nu}\delta^i_0.$  For the line element (\ref{1}), the independent set of Einstein field equations are then obtained as
\begin{eqnarray}
\label{3} \rho &=& \frac{1}{r^{2}} \left[ r(1-e^{-2\lambda})
\right]',\\
\label{4} p_r &=& - \frac{1}{r^{2}} \left( 1-e^{-2\lambda} \right)
+
\frac{2\nu'}{r}e^{-2\lambda} ,\\
\label{5}p_t &=& e^{-2\lambda}\left( \nu'' + \nu'^{2} +
\frac{\nu'}{r}- \nu'\lambda' - \frac{\lambda'}{r} \right) ,
\end{eqnarray}
where a prime ($'$) denotes differentiation with respect to $r$. In the field equations (\ref{3})-(\ref{5}), we have assumed $8\pi G=1=c$. The system of equations determines the behaviour of the gravitational field of an anisotropic imperfect fluid sphere. The mass contained within a radius $r$ of the sphere is defined as
\begin{equation}
\label{6} m(r)= \frac{1}{2}\int_0^r\omega^2 \rho(\omega)d\omega.
\end{equation}
A different but equivalent form of the field equations can be obtained if we introduce the transformation\cite{Durga}
\begin{equation}
\label{7} x = Cr^2,~~ Z(x)  = e^{-2\lambda(r)} ~\mbox{and}~
A^{2}y^{2}(x) = e^{2\nu(r)},
\end{equation}
where $A$ and $C$ are arbitrary constants. Under the transformation (\ref{7}), the system of equations (\ref{3})-(\ref{5}) take the following form
\begin{eqnarray}
\label{9}   \frac{\rho}{C} &=& \frac{1-Z}{x} - 2\dot{Z}   , \\
\label{10} \frac{p_r}{C} &=& 4Z\frac{\dot{y}}{y} + \frac{Z-1}{x}
, \\
\label{11} p_t &=& p_r +\Delta \\
\label{12}0 & =&\dot{Z} \left(2x^2\dot{y}+xy\right)+
Z(4x^2\ddot{y}-y)\nonumber\\
&&+\left(1-\frac{\Delta x}{C}\right)y,
\end{eqnarray}
where $\Delta = p_t-p_r$ is the measure of anisotropy and dots denote differentiation with respect to the variable $x$. The anisotropic stress will be directed outward (repulsive) when $p_t > p_r$ (i.e., $ \Delta > 0$) and inwards when $p_t < p_r$ (i.e.,  $\Delta < 0$). 

\section{\label{sec3} Method of generating  analytic solutions}
The system (\ref{9})-(\ref{12}) comprises four equations in the six unknowns namely,  $Z, y, \rho, p_r, p_t$ and $\Delta$.
Therefore, we have the freedom to choose any two variables to integrate the system. In our formalism, rather than assuming an equation of state (EOS) for the matter composition, we assume a particular form of $y$ together with a prescribed anisotropy $\Delta$ which are well-behaved and can provide solutions to Eq.~(\ref{12}). We choose the the metric function $y$  as
\begin{equation}
\label{eq:b5} y = (1+ a x^n)^m
\end{equation}
where $a, m$ and $n$ are real numbers. Substitution of Eq.~(\ref{eq:b5}) in (\ref{12}) yields
\begin{eqnarray}
\label{eq:b6}&&\dot{Z}+f(x) Z -\frac{\left(\frac{\Delta x}{C}-1\right)(1+ax^n)}{x[1+(2mn+1)ax^n]}=0,\\
&&f(x)= \frac{1}{x(1+ax^n)[1+(2mn+1)ax^n]}\nonumber\\
&&\times\left[(4mn(mn-1)-1)(ax^n)^{2}\right.\nonumber\\
&&\left.+(4mn(n-1)-2)ax^n -1\right].\nonumber
\end{eqnarray}
Using partial fractions, we write Eq.~(\ref{eq:b6}) in the form
\begin{eqnarray}
\label{eq:b7}&& \dot{Z} + g(x) Z = \frac{\left(\frac{\Delta x}{C}-1\right)(1+ax^n)}{x[1+(2mn+1)ax^n]},\\
&&g(x)= \left[- \frac{1}{x} +
\frac{2n(m-1)ax^{n-1}}{(1+ax^n)} \right.\nonumber\\
&&\left. +\frac{2n[2m(n-1)+1]ax^{n-1}}{[1+(2mn+1)ax^n]} \right],\nonumber
\end{eqnarray}
whose solution can be expressed in integral form as
\begin{eqnarray}
 \label{eq:b8}Z &=& \frac{x}{(1+ax^n)^{2(m-1)}[1+
 (2mn+1)ax^n]^{\frac{2[2m(n-1)+1]}{2mn+1}}} \nonumber\\
 && \times \left[
 \int \left(\frac{\Delta}{Cx}-\frac{1}{x^2}\right) (1+ax^n)^{(2m-1)}\right.\nonumber\\
&&\left. \times[1+(2mn+1)ax^n]^{1- \frac{4m}{2mn+1}} dx - B
 \right],\nonumber\\
\end{eqnarray}
where $B$ is the constant of integration. At this stage, we specify the anisotropy $\Delta$. We assume the radial fall-off profile of the anisotropic parameter in the form
\begin{equation}
\label{eq:e}\Delta=\frac{\alpha Ca
x}{(1+ax^n)^{(2m+1)}[1+(2mn+1)ax^n]^{1- \frac{4m}{2mn+1}}},
\end{equation}
where the constant $\alpha$ specifies the extend of anisotropy. This particular choice ensures that anisotropy vanishes at the center of the star. Substitution of (\ref{eq:e}) into (\ref{eq:b8}) yields
\begin{eqnarray}
 \label{eq:e1}Z &=& \frac{x}{(1+ax^n)^{2(m-1)}[1+
 (2mn+1)ax^n]^{\frac{2[2m(n-1)+1]}{2mn+1}}} \nonumber\\
 && \times \left[\int \left(\frac{\alpha}{(1+ax^n)^2} -
   l(x)\right) dx - B \right],\\
l(x)&=& \frac{(1+ax^n)^{(2m-1)}[1+(2mn+1)ax^n]^{1- \frac{4m}{2mn+1}}}{x^2},\nonumber
\end{eqnarray}
which solves the system. It should be mentioned here that in an earlier work Herrera {\em et al}\cite{Herrera08} had shown that all static spherically symmetric anisotropic solutions to Einstein field equations could be obtained by making use of two generating functions. In our case, it turns out that the solution (\ref{eq:e1}) can be obtained as a special case by choosing the
following generating functions
\begin{eqnarray}
z(r) &=& \frac{1}{r}+\frac{2m n a C^n r^{2n-1}}{1+a C^n r^{2n}};\nonumber\\
\Pi (r) &=& \frac{-8 \pi\alpha C a x}{(1+a x^n)^{(2m+1)}[1+(2m n+1)a x^n]^{1- \frac{4m}{2m n+1}}}
\nonumber
\end{eqnarray} 
in equation ($10$) of Ref.~\cite{Herrera08}.

We are now in a position to integrate equation (\ref{eq:e1}) for specified values of $m$ and $n$. Interestingly, it turns out that the solutions can also be expressed in terms of elementary functions for particular values of $m$ and $n$ as will be shown in the following section.

\section{\label{sec4} Anisotropic models}
It is interesting to note that an anisotropic generalization of a large family of physically reasonable isotropic stellar models studied earlier can be regained by suitably fixing the values of $m$ and $n$. Our motivation will be to generate new solutions only for those values of $m$ and $n$ which would allow us to regain the isotropic analogues of solutions which have been shown to be well-behaved and physically acceptable\cite{Delgaty98}. It is remarkable that the new class of solutions, as shown below, contains a large class of known solutions which have been developed to study relativistic isotropic fluid spheres.  

\subsection{\bf Case I: $m=\frac{1}{2}$ and $n=1$:}
By setting $m=\frac{1}{2}$ and $n=1$ in Eqs.~(\ref{eq:e1}) and (\ref{eq:e}), we obtain
\begin{eqnarray}
\label{eq:b9} Z&=&\frac{(1+ax)(1-Bx)}{(1+2ax)}-\frac{\alpha x}{(1+2ax)},\\
\Delta&=&\frac{\alpha  C ax}{(1+ax)^2},
\end{eqnarray}
so that the line element (\ref{1}) takes the form
\begin{eqnarray}
 ds^{2} &=& -A^2(1+aCr^2) dt^{2} +\nonumber\\
&& \left[\frac{(1+aCr^2)(1-BCr^2)}{(1+2aCr^2)}-\frac{\alpha Cr^2}{(1+2aCr^2)}\right]^{-1}dr^{2}
\nonumber\\
&& \label{eq:b10}+
 r^{2}(d\theta^{2} + \sin^{2}{\theta} d\phi^{2}).
\end{eqnarray}

\subsubsection{Anisotropic generalization of Tolman IV Model}
In (\ref{eq:b10}), if we set $a=\frac{1}{D^2}, B=\frac{1}{R^2}$ and $C=1$, the line element (\ref{eq:b10}) reduces to
\begin{eqnarray}
ds^{2}& =& -A^2\left(1+\frac{r^2}{D^2}\right) dt^{2}\nonumber\\
&&+\left[\frac{\left(1+\frac{r^2}{D^2}\right)
\left(1- \frac{r^2}{R^2}\right)}{\left(1+2\frac{r^2}{D^2}\right)}
-\frac{\alpha r^2}{\left(1+2\frac{r^2}{D^2}\right)}\right]^{-1}dr^{2} \nonumber\\
&& \label{eq:b11} +
 r^{2}(d\theta^{2} + \sin^{2}{\theta} d\phi^{2}).
\end{eqnarray}
Note that for zero anisotropy $\Delta=0$ (i.e., $\alpha =0$), the line element (\ref{eq:b11}) reduces to
\begin{eqnarray}
\label{eq:e2} ds^{2} &=& -A^2\left(1+\frac{r^2}{R^2}\right) dt^{2} +
\frac{1+2\frac{r^2}{D^2}}{\left(1+\frac{r^2}{D^2}\right)\left(1-\frac{r^2}{R^2}\right)}dr^{2}\nonumber\\
&&+ r^{2}(d\theta^{2} + \sin^{2}{\theta} d\phi^{2}),
\end{eqnarray}
which is the well known Tolman IV solution\cite{Tolman}. Thus, the metric (\ref{eq:b11}) turns out to be an anisotropic generalization of Tolman IV solution. This solution was shown to satisfy all the physical requirements of a realistic star \cite{Delgaty98} and previously used by Tolman\cite{Tolman} to study relativistic compact stars with isotropic matter distribution. It is to be noted that an anisotropic generalization of the Tolman IV solution was obtained earlier by Cosenza {\em et al} \cite{Cosenza}. While in the earlier approach the generalization was done by assuming a specific density profile (or equivalently making an ansatz for the metric potential $g_{rr}$), in our case, the metric potential $g_{rr}$ gets determined for a specific form of the metric potential $g_{tt}$.

\subsubsection{Anisotropic generalization of  de-Sitter solution}
For $a=-1, A=1, B=2$ and $C=\frac{1}{R^2}$, the line element (\ref{eq:b10}) becomes
\begin{eqnarray}
 ds^{2} &=& -\left(1- \frac{r^2}{R^2} \right) dt^{2}
+
\left[\left(1-\frac{r^2}{R^2}\right)-\frac{\alpha \frac{r^2}{R^2}}{\left(1-2\frac{r^2}{R^2}\right)}\right]^{-1}dr^{2} \nonumber\\
&& \label{eq:e3} +
 r^{2}(d\theta^{2} + \sin^{2}{\theta} d\phi^{2}).
\end{eqnarray}
Now, if we set $\alpha =0$ (i.e., $\Delta=0$), the metric reduces to the familiar de-Sitter solution
\begin{eqnarray}
\label{eq:b12} ds^{2} &=& -\left(1-\frac{r^2}{R^2}\right) dt^{2} +
\left(1-\frac{r^2}{R^2}\right)^{-1}dr^{2} \nonumber\\
&&+ r^{2}(d\theta^{2} + \sin^{2}{\theta} d\phi^{2}),
\end{eqnarray}
which models an isotropic universe dominated by dark energy which is, in general, interpreted in terms of a cosmological constant.

\subsubsection{Anisotropic generalization of Einstein universe}
For $a=0, B=1$ and $C=\frac{1}{R^2}$, using Eq.~(\ref{eq:b10}), we obtain
\begin{eqnarray}
\label{eq:b13} ds^{2} &=& -A^2 dt^{2} + \left(1-\frac{r^2}{R^2}-
\alpha \frac{r^2}{R^2}\right)^{-1}dr^{2}\nonumber\\
&& + r^{2}(d\theta^{2} + \sin^{2}{\theta} d\phi^{2}).
\end{eqnarray}
For vanishing anisotropy ($\alpha =0$), the metric reduces to the isotropic Einstein universe model described by the metric 
\begin{equation}
\label{eq:e4} ds^{2} = -A^2 dt^{2} +
\left(1-\frac{r^2}{R^2}\right)^{-1}dr^{2} +
 r^{2}(d\theta^{2} + \sin^{2}{\theta} d\phi^{2}).
\end{equation}
In this case, the metric corresponds to a matter dominated Friedmann model with zero curvature in which the universe will continue to expand forever with the right amount of energy provided during the time of the Big-bang.

\subsection{\bf Case II: $m=n=1$:}
\subsubsection{Anisotropic generalization of Korkina and Orlyanskii solution III}
By setting $m=n=1$ and using Eqs.~(\ref{eq:e1}) and (\ref{eq:e}), we obtain
\begin{eqnarray}
\label{eq:a17} Z &=&1-Bx(1+3ax)^{-2/3} \nonumber\\
&&-\alpha x (1+ax)^{-1}(1+3ax)^{-2/3},\\
\Delta &=&\frac{\alpha Cax(1+3ax)^{1/3}}{(1+ax)^3},
\end{eqnarray}
so that the line element (\ref{1}) takes the form
\begin{eqnarray}
\label{eq:a18} ds^{2} &=& -A^2(1+aCr^2)^2 dt^{2} + \left[1-BCr^2(1+3aCr^2)^{-2/3} \right.\nonumber\\
&&\left.  -\alpha Cr^2(1+aCr^2)^{-1}(1+3aCr^2)^{-2/3}  \right]^{-1}dr^{2}\nonumber\\
&&+ r^{2}(d\theta^{2} + \sin^{2}{\theta} d\phi^{2}).
\end{eqnarray}
Now, for an anisotropic sphere ($\alpha=0$), if we set $B=0$ and C=1, the metric (\ref{eq:a18}) reduces to 
\begin{equation}
\label{eq:a19} ds^{2} = -A^2(1+ar^2)^2 dt^{2}  + dr^{2} +
r^{2}(d\theta^{2} + \sin^{2}{\theta} d\phi^{2}),
\end{equation}
which is the Korkina and Orlyanskii solution III\cite{Korkina}. Consequently, the metric (\ref{eq:a18}) is a generalization of the solution of Korkina and Orlyanskii\cite{Korkina}.

\subsection{\bf Case III: $m=\frac{3}{2}$ and $n=1$:}
\subsubsection{Anisotropic generalization of Heintzmann IIa  solution}
When $m=\frac{3}{2}$ and $n=1$, using Eqs.~(\ref{eq:e1}) and (\ref{eq:e}), we obtain
\begin{eqnarray}
\label{eq:b14} Z&=&\frac{2-ax- 2Bx(1+4ax)^{-1/2}}{2(1+ax)} \nonumber\\
&&- \frac{\alpha x(1+4ax)^{-1/2}}{(1+ax)^2},\\
\Delta&=&\frac{\alpha Cax\sqrt{1+4ax}}{(1+ax)^4}.
\end{eqnarray}
Now, if we set $B=3a C/2$, Eq.~(\ref{eq:b14}) takes the form
\begin{equation}
\label{eq:b15} Z=1-\frac{3ax }{2}
\left[\frac{1+c(1+4ax)^{-1/2}}{1+ax}\right]- \frac{\alpha
x(1+4ax)^{-1/2}}{(1+ax)^2},
\end{equation}
and consequently the metric (\ref{1}) gets the form
\begin{eqnarray}
\label{eq:e5} ds^{2} &=& -A^2(1+ar^2)^3 dt^{2} +
\left(1-\frac{3ar^2 }{2}\right.\nonumber\\
&&\left.\times\left[\frac{1+c(1+4ar^2)^{-1/2}}{1+ar^2}\right]- \frac{\alpha
r^2(1+4ar^2)^{-1/2}}{(1+ar^2)^2}\right)^{-1}dr^{2}
\nonumber\\
&& + r^{2}(d\theta^{2} + \sin^{2}{\theta} d\phi^{2}),
\end{eqnarray}
where we have set $C=1$.  The above metric reduces to Heintzmann IIa\cite{Heintzmann} solution
\begin{eqnarray}
\label{eq:b16} ds^{2} &=& -A^2(1+ar^2)^3 dt^{2} +\nonumber\\
&&\left(1-\frac{3ar^2 }{2}
\left[\frac{1+c(1+4ar^2)^{-1/2}}{1+ar^2}\right]\right)^{-1}dr^{2}
\nonumber\\
&& + r^{2}(d\theta^{2} + \sin^{2}{\theta} d\phi^{2}),
\end{eqnarray}
when $\alpha=0$.

\subsection{\bf Case IV: $m=2$ and $n=1$:}
\subsubsection{Anisotropic generalization of Durgapal IV  Model}
For $m=2$ and $n=1$, using Eqs.~(\ref{eq:e1}) and (\ref{eq:e}), we obtain
\begin{eqnarray}
\label{eq:b25}
Z&=&\frac{7-10ax-a^2x^2}{7(1+ax)^2}-\frac{Bx}{(1+ax)^2(1+5ax)^{2/5}} \nonumber\\
&&-\frac{\alpha x}{(1+ax)^3(1+5ax)^{2/5}},\\
\Delta&=&\frac{\alpha Cax(1+5ax)^{3/5}}{(1+ax)^5},
\end{eqnarray}
and consequently, the line element (\ref{1}) takes the form
\begin{eqnarray}
\label{eq:b26} ds^{2}& =& -A^2(1+aCr^2)^4 dt^{2} +
\left[\frac{7-10aCr^2-a^2C^2r^4}{7(1+aCr^2)^2}\right.\nonumber\\
&&\left.-\frac{BCr^2}{(1+aCr^2)^2(1+5aCr^2)^{2/5}}\right.\nonumber\\
&&\left.-\frac{\alpha
Cr^2}{(1+aCr^2)^3(1+5aCr^2)^{2/5}}\right]^{-1}dr^{2}\nonumber\\
&& + r^{2}(d\theta^{2} + \sin^{2}{\theta} d\phi^{2}).
\end{eqnarray}
Now, if we set $a=1$ and $\alpha =0$, we regain the Durgapal IV\cite{Durga82} metric
\begin{eqnarray}
\label{eq:e6} ds^{2}& =& -A^2(1+Cr^2)^4 dt^{2} +
\left[\frac{7-10Cr^2-C^2r^4}{7(1+Cr^2)^2}\right.\nonumber\\
&&\left.-\frac{BCr^2}{(1+Cr^2)^2(1+5Cr^2)^{2/5}}\right]^{-1}dr^{2}
\nonumber\\
&& + r^{2}(d\theta^{2} + \sin^{2}{\theta} d\phi^{2}).
\end{eqnarray}

\subsection{\bf Case V: $m=\frac{5}{2}$ and $n=1$:}
\subsubsection{Anisotropic generalization of Durgapal V  Model}
For $m=\frac{5}{2}$ and $n=1$, using Eqs.~(\ref{eq:e1}) and (\ref{eq:e}) we obtain
\begin{eqnarray}
\label{eq:b27} Z&=&\frac{1-\frac{ax(309 +54
ax+8a^2x^2)}{112}-\frac{Bx}{(1+6ax)^{1/3}}}{(1+ax)^3} \nonumber\\
&&- \frac{\alpha x}{(1+ax)^4(1+6ax)^{1/3}},\\
\Delta&=&\frac{\alpha Cax(1+6ax)^{2/3}}{(1+ax)^6},
\end{eqnarray}
and subsequently the line element (\ref{1}) takes the form
\begin{eqnarray}
\label{eq:b28} ds^{2} &=& -A^2(1+aCr^2)^5 dt^{2} +\nonumber\\
&&\left[\frac{1-\frac{aCr^2(309 +54a
Cr^2+8a^2C^2r^4)}{112}-\frac{BCr^2}{(1+6aCr^2)^{1/3}}}{(1+aCr^2)^3}
\right. \nonumber\\
&& \left. - \frac{\alpha Cr^2}{(1+aCr^2)^4(1+6aCr^2)^{1/3}}
\right]^{-1}dr^{2}\nonumber\\
&& + r^{2}(d\theta^{2} + \sin^{2}{\theta} d\phi^{2}).
\end{eqnarray}
Obviously, the line element ({\ref{eq:b28}})  reduces to Durgapal V\cite{Durga82} solution if we set $a=1$ and $\alpha =0$.

\subsection{\bf Case VI: $m=\frac{1}{4}$ and $n=1$:}
\subsubsection{Anisotropic generalization of Durgapal {\em et al}\cite{Durga84} stellar model}
For $m=-\frac{1}{4}$ and $n=1$, using Eqs.~(\ref{eq:e1}) and (\ref{eq:e}), we obtain
\begin{eqnarray}
\label{eq:b29}
Z&=&\frac{u(x)}{(2+ax)^4},\\
\Delta&=&\frac{8\alpha Cax }{\sqrt{1+ax}(2+ax)^3},\\
u(x)&=&-16x(1+ax)^{3/2}(B(1+ax)+\alpha)\nonumber\\
&&+4(1+ax)^2(4+4ax-a^2x^2),\nonumber
\end{eqnarray}
so that the line element (\ref{1}) takes the form
\begin{eqnarray}
\label{eq:b30} ds^{2} &=& -\frac{A^2}{\sqrt{1+aCr^2}} dt^{2}
\nonumber\\
&& + \frac{(2+aCr^2)^4}{v(x)}dr^{2}
\nonumber\\
&& + r^{2}(d\theta^{2} + \sin^{2}{\theta} d\phi^{2}),\\
v(x) &=&-16Cr^2(B(1+aCr^2)+\alpha
)(1+aCr^2)^{3/2}\nonumber\\
&&+4(1+aCr^2)^2(4+4aCr^2-a^2C^2r^4).\nonumber
\end{eqnarray}
By setting $a=-1$ and $\alpha =0$, the above metric can be reduced to the Durgapal {\em et al}\cite{Durga84} stellar model.

\section{\label{sec5} Physical acceptability}
\label{sec:4}
In the previous section, we have presented an algorithm to generate a large class of anisotropic solutions and showed that many well-known exact solutions may be regained by suitably fixing the model parameters in this formulation. To check physical acceptability of our class of solutions, we consider a particular solution (case I). The variables in this case are obtained as
\begin{eqnarray}
e^{2\nu} &=&A^2(1+ax), \\
e^{2\lambda} &=& \frac{(1+2ax)}{(1+ax)(1-Bx)-\alpha x},\\
\frac{\rho}{C} &=&
\frac{(a+\alpha)(3+2ax)+B[3+ax(7+6ax)]}{(1+2ax)^2},\\
\frac{p_r}{C} &=& \frac{a(1+ax)-\alpha (1+3ax)-
B[1+ax(4+3ax)]}{(1+ax)(1+2ax)},\\
p_t &=& p_r+\Delta, \\
\frac{\Delta}{C} &=& \frac{\alpha ax}{(1+ax)^2}.
\end{eqnarray}
The physical quantities are expressed in simple elementary functions which facilitates a detailed study of the physical behaviour of the star. Most importantly, the solution contains an `anisotropic switch' $\alpha$ which can be conveniently used to investigate the impact of anisotropy.
Another interesting feature of our solution is that the solution provides a barotropic equation of state (EOS) $p_r = p_r(\rho)$ which is obtained explicitly in the form
\begin{eqnarray} \label{state}
\frac{p_r}{C} &=& \frac{1}{8} \left[\frac{24\alpha
(3B+2\tilde{\rho})+\beta (a-7\alpha -2B- \tilde{\rho})}{a+\alpha
-2B -\tilde{\rho}}\right.\nonumber\\
&&\left.+\frac{2B\beta}{2(a+\alpha)-B}+ 14\alpha
-2a-13B \right],
\end{eqnarray}
where we have used the relation
\[\tilde{\rho}=\frac{\rho}{C}~~\mbox{and}~~\beta
=\sqrt{[2(a+\alpha)-B][2(a+\alpha)-23B+16 \tilde{\rho}]}.\] 
We would like to stress here that a barotropic EOS is generally difficult to extract from an exact solution of field equations. It is not so in our case.

Let us now analyze the physical acceptability of our solution:
\begin{enumerate}
\item[i.]
In our model, we have $(e^{2\nu(r)})'_{r=0}=(e^{2\lambda(r)})'_{r=0}=0 $
and  $e^{2\nu (0)}=A^2,~ e^{2\lambda(0)}=1$; these imply that the metric is regular at the centre $r=0$.

\item[ii.]  Since $\rho(0)= 3C(a+B+\alpha)$ and $\displaystyle p_r(0)=p_t(0)=C (a-B- \alpha)$, the energy density, radial pressure and tangential pressure will be non-negative at the centre if we choose the parameters satisfying the condition $a>B+\alpha$.

\item[iii.]
The condition $p_r(r=s)=0$ determines the boundary of the star
$$s=\sqrt{\frac{a-(4B+3\alpha)+\sqrt{(a+2B)^2-6
(a-2B)\alpha +9 {\alpha}^2}}{6aBC}}.$$

\item[iv.] The interior solution (\ref{eq:b10}) should be matched to the exterior  Schwarzschild metric
\begin{eqnarray}
 \label{eq:exterior}
ds^2 &=& - \left( 1 - \frac{2M}{r} \right) dt^2 +\left( 1 -
\frac{2M}{r} \right)^{-1} dr^2 \nonumber\\
&& + r^2 (d\theta^2 +\sin^2\theta d\phi^2),
\end{eqnarray}
across the boundary boundary of the star $r=s$, where $M$ is the total mass of the sphere which can be obtained directly from Eq.~(\ref{6}) as
$$\displaystyle M=m(s)=\frac{Cs^3[a+\alpha + B(1+aCs^2)]}{2(1+2aCs^2)}.$$
Matching of the line elements (\ref{eq:b10}) and (\ref{eq:exterior}) at the boundary $r=s$ yields
\begin{eqnarray}
 \label{eq:f43}\left( 1 - \frac{2M}{s}
\right)&=&\frac{(1+aCs^2)(1-BCs^2)}{(1+2aCs^2)}\nonumber\\
&&-\frac{\alpha Cs^2}{(1+2aCs^2)},\\
\label{eq:f44}\left( 1 - \frac{2M}{s} \right)&=&A^2(1+aCs^2).
\end{eqnarray}
Making use of the junction conditions, the constant $A$ is determined as
$$\displaystyle A^2= \frac{\sqrt{(a+2B)^2-6
(a-2B)\alpha +9 {\alpha}^2}+ 2B+3 \alpha -a}{4a}.$$

\item[v.] The gradient of density, radial pressure and tangential pressure are respectively obtained as
\begin{eqnarray}
\frac{d\rho}{dr}&=& - \frac{2ac^2[2(a+\alpha)+B]r(5+2aCr^2)}{(1+2aCr^2)^3},\\
\frac{dp_r}{dr}&=&\frac{2 a C^2 r}{(1+a C r^2)^2 (1+2 a Cr^2)^2}\nonumber\\
&&\times \left[-(2 a+B)(1+a C r^2)^2\right.\nonumber\\
&&\left. + 2 a C r^2(2+3 a C r^2)\alpha\right],\\
\frac{dp_t}{dr}&=&\frac{2 a C^2 r}{(1+a C r^2)^2 (1+2 a Cr^2)^2}\nonumber\\
&&\times\left[-(2 a+B)(1+a C r^2)^3+\right.\nonumber\\
&&\left. (1+ a C r^2(7+2 a C r^2(5+a C r^2)))\alpha\right].
\end{eqnarray}
The decreasing nature of these quantities is shown graphically.

\item[vi.] Within a stellar interior, it is expected that the speed of sound should be less than the speed of light i.e., $\displaystyle 0
\leq \frac{dp_r}{d\rho}\leq 1$ and $\displaystyle 0 \leq
\frac{dp_t}{d\rho}\leq 1$.\\
In our model, we have
\begin{eqnarray}
\frac{dp_r}{d\rho}&=&\frac{1}{(1+a C r^2)^2 (5+2 a C r^2) (2 a+B+2 \alpha )}\nonumber\\
&&\times(1+2 a C r^2)\left[(2 a+B)(1+a C r^2)^2\right.\nonumber\\
&&\left. -2 a C r^2(2+3 a C r^2)\alpha\right],\\
\frac{dp_t}{d\rho}&=&\frac{1}{(1+a C r^2)^3 (5+2 a C r^2) (2 a+B+2 \alpha )}\nonumber\\
&&\times(1+2 a C r^2)\left[(2 a+B)(1+a C r^2)^3\right.\nonumber\\
&&\left.-(a C r^2(7+2 a C r^2(5+a C r^2)))\alpha\right].
\end{eqnarray}
By choosing the model parameters appropriately, we have shown that this requirement is also satisfied in our model.

\item[vii.] The fulfillment of energy conditions for an anisotropic fluid i.e., $\rho -p_r-2p_t\geq0$ and $\rho +p_r+2p_t\geq 0$ are also shown to be satisfied in this model.

\item[viii.] Finally, we have calculated the adiabatic index  
\begin{equation}
\Gamma=\frac{\rho+p}{p}\frac{dp}{d\rho},\nonumber
\end{equation} 
for a particular configuration. Bondi's \cite{Bondi64} analyses  show that a Newtonian isotropic sphere will be in  equilibrium if the adiabatic index $\Gamma > 4/3$ which, however, gets modified for a relativistic anisotropic fluid sphere. Subsequently, the issue relating to stability of a relativistic anisotropic spherical body was taken up by many investigators (see for example \cite{Herrera79,Chan93,Esculpi07} and references therein). Based on these results, it can be concluded that an anisotropic fluid sphere will be stable if the following condition is fulfilled: 
\begin{equation}
\Gamma > \frac{4}{3} - \left[ \frac{4}{3}\frac{(p_r-p_t)}{|{p_r}'|r}\right]_{max}\label{eq_stab}.
\end{equation}
In Fig.~\ref{fig8} and \ref{fig9}, we have plotted $\Gamma$ for $\alpha=0.5$ and $\alpha=0$, respectively. The dashed lines in the plots correspond to values of the right hand side of equation (\ref{eq_stab}). We note that in the case of $\alpha=0.5$, the term within bracket $[~]$ takes its maximum value at the centre and decreases radially outward. Consequently, the above condition is satisfied throughout the star. In the case of an isotropic configuration ($\alpha=0$), the right hand side of equation (\ref{eq_stab}) remains constant ($4/3$) throughout the star and it is obvious from Fig.~\ref{fig9} that the stability requirement is fulfilled in this case as well.  

\end{enumerate}

\section{\label{sec6} Compatibility with observational data}
We examine the physical applicability of our solution by making use of the values of masses and radii of observed pulsars as input parameters.  To illustrate the case, we have considered the data available from the pulsar $4U1820-30$ whose estimated mass and radius are $M = 1.58 ~M_{\odot}$ and $s = 9.1~km$, respectively (\cite{Guver10}). For these values, we have determined two sets of constants. For the isotropic case ($\alpha=0$), we have obtained $A=0.7375,~C=0.0068,~B=0.2719$ and assuming the star to be composed of anisotropic matter (we have assume $\alpha =0.5$) the constants have been evaluated as $A=0.7375,~C=0.0068,~B=-0.1188$. Note that the parameter $a$ remains free in this model. Making use of these values, we have shown graphically the nature of all the physically meaningful quantities in Fig.~(\ref{fig1})-(\ref{fig9}). The plots show that all the physically meaningful variables comply with the requirements of a realistic star. In particular, the figures highlight the effects of anisotropy on the gross physical behaviour of a compact star.

\begin{figure}
\resizebox{0.5\textwidth}{!}{
  \includegraphics{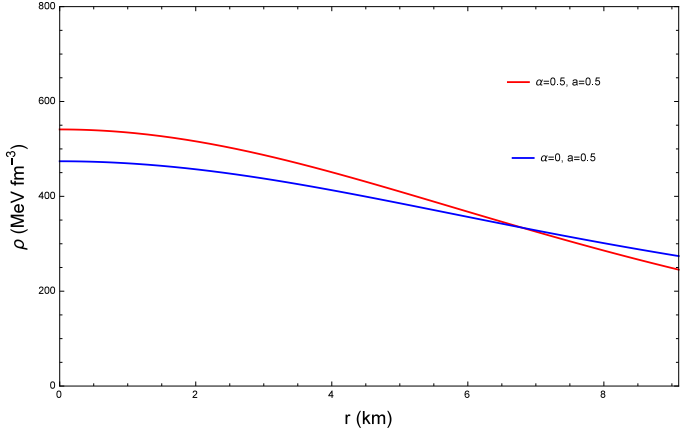}
}
\caption{Density profiles.}
\label{fig1}
\end{figure}

\begin{figure}
\resizebox{0.5\textwidth}{!}{
  \includegraphics{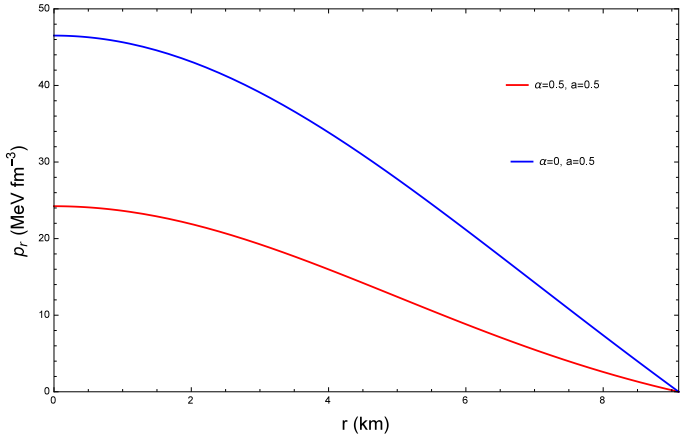}
}
\caption{Radial pressure profiles.}
\label{fig2}
\end{figure}

\begin{figure}
\resizebox{0.5\textwidth}{!}{
  \includegraphics{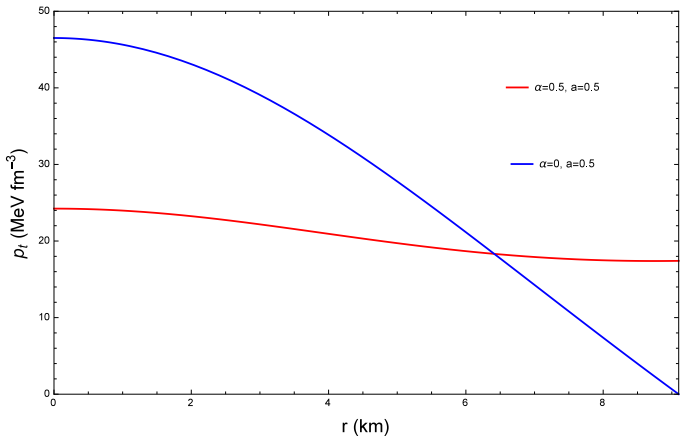}
}
\caption{Tangential pressure profiles.}
\label{fig3}
\end{figure}

\begin{figure}
\resizebox{0.5\textwidth}{!}{
  \includegraphics{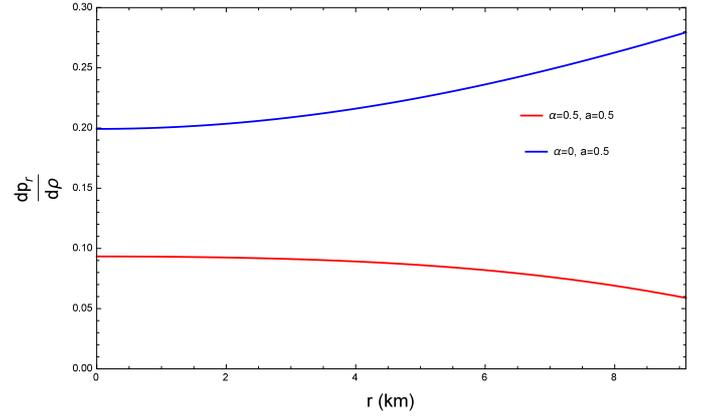}
}
\caption{Radial component of sound speed.}
\label{fig4}
\end{figure}

\begin{figure}
\resizebox{0.5\textwidth}{!}{
  \includegraphics{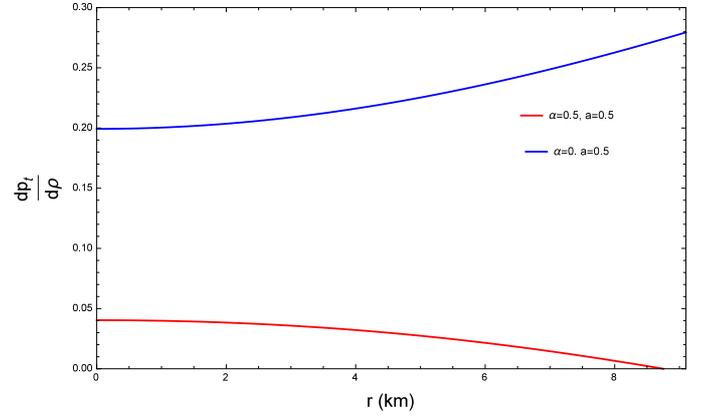}
}
\caption{Transverse component of sound speed.}
\label{fig5}
\end{figure}

\begin{figure}
\resizebox{0.5\textwidth}{!}{
  \includegraphics{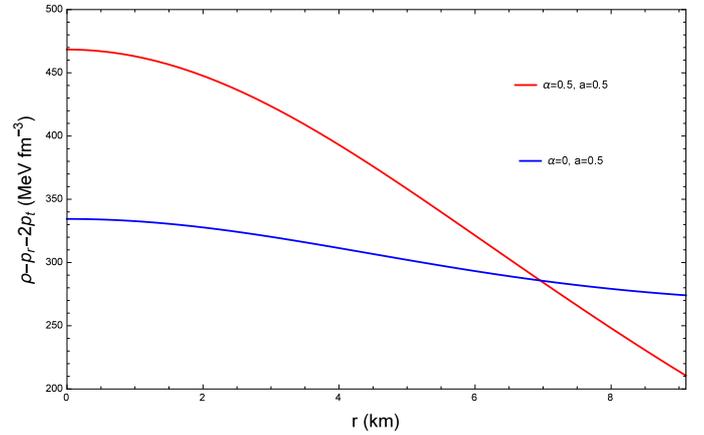}
}
\caption{Fulfillment of energy condition.}
\label{fig6}
\end{figure}

\begin{figure}
\resizebox{0.5\textwidth}{!}{
  \includegraphics{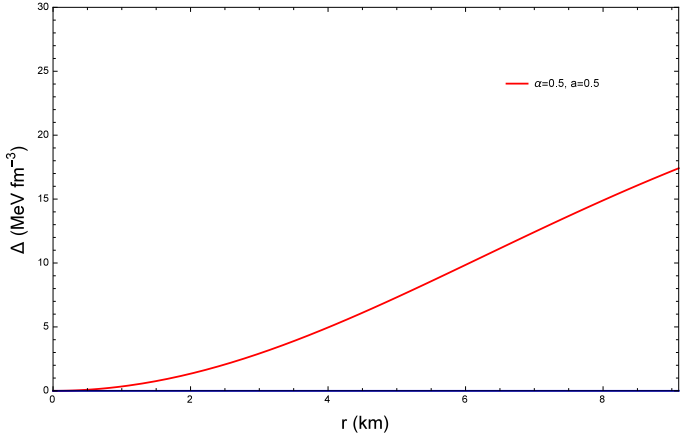}
}
\caption{Radial variation of anisotropy.}
\label{fig7}
\end{figure}

\begin{figure}
\resizebox{0.5\textwidth}{!}{
  \includegraphics{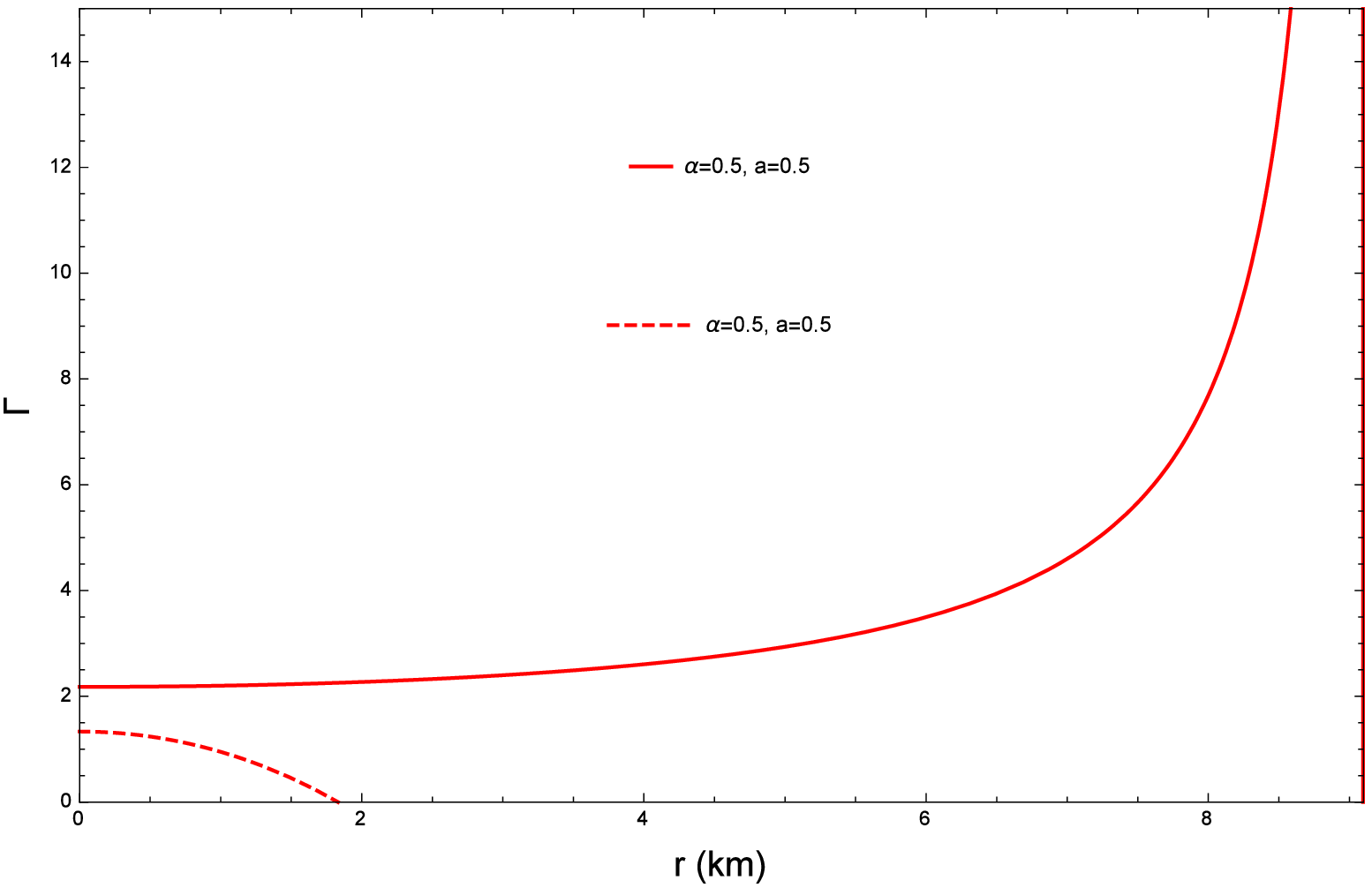}
}
\caption{Radial variation of adiabatic index for $\alpha \neq 0$. The dashed line corresponds to the right hand side of Eq.~\ref{eq_stab}.}
\label{fig8}
\end{figure}

\begin{figure}
\resizebox{0.5\textwidth}{!}{
  \includegraphics{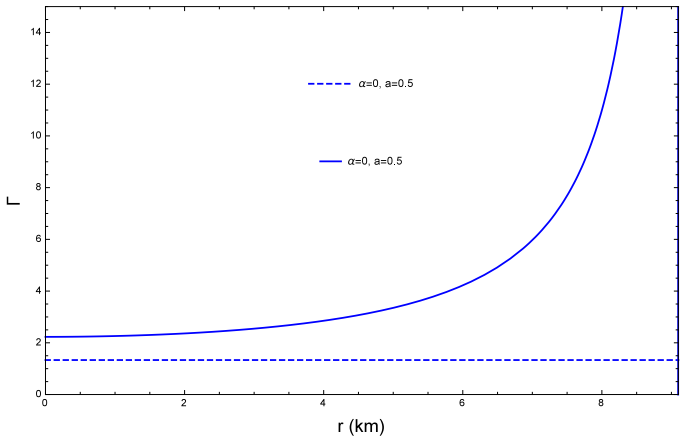}
}
\caption{Radial variation of adiabatic index for $\alpha = 0$. The dashed line corresponds to the right hand side of Eq.~\ref{eq_stab}.}
\label{fig9}
\end{figure}

\section{\label{sec7} Conclusions}
To summarize, we have developed an algorithm to generate exact solutions to Einstein field equations for a spherically symmetric anisotropic star. The most remarkable feature of our approach is that a large family of previously developed isotropic stellar solutions can be regained from our anisotropic family of solutions by suitably fixing the model parameters in our treatment. It will be interesting to explore the possibility of generating new class of solutions by choosing sets of values of $m$ and $n$ which have not been considered in this work. Probing the effects of electromagnetic field on top of anisotropy is another area which we would also like to take up in our future investigation and will be reported elsewhere. 

\section{Acknowledgements}
\noindent The work of RS is supported by the MRP grant F.PSW-195/15-16 (ERO) of the UGC, Govt. of India. RS also gratefully acknowledges support from the Inter-University Centre for Astronomy and Astrophysics (IUCAA), Pune, India, under its Visiting Research Associateship Programme. 

The authors are grateful to the anonymous referee for his helpful comments and suggestions.

\end{document}